# Impurity conduction in phosphorus-doped buried-channel silicon-on-insulator field-effect transistors

Yukinori Ono*, Jean-Francois Morizur, Katsuhiko Nishiguchi, Kei Takashina,

Hiroshi Yamaguchi,

*NTT Basic Research Laboratories, NTT Corporation, 3-1 Morinosato Wakamiya,*

*Atsugi, Kanagawa 243-0198, Japan*

Kazuma Hiratsuka, Seiji Horiguchi

*Department of Electrical and Electronic Engineering, Faculty of Engineering and Resource Science,*

*Akita University 1-1 Tegata-gakuen-machi, Akita-shi, Akita, 010-8502 Japan*

Hiroshi Inokawa,

*Research Institute of Electronics, Shizuoka University, 3-5-1, Johoku, Hamamatsu, 432-8011, Japan*

Yasuo Takahashi

*Graduate School of Information Science and Technology, Hokkaido University, Sapporo,*

*Hokkaido 060-0814, Japan*

**Corresponding author**

*E*-mail address: ono@aecl.ntt.co.jp

Tel: +81-46-240-2641

Fax:+81-46-240-4317

**Abstract:**

We investigate transport in phosphorus-doped buried-channel metal-oxide-semiconductor field-effect transistors at temperatures between 10 and 295 K. We focus on transistors with phosphorus donor concentrations higher than those previously studied, where we expect conduction to rely on donor electrons rather than conduction-band electrons. In a range of doping concentration between around 2.1 and 8.7 x $10^{17}$ $cm^{-3}$, we find that a clear peak emerges in the conductance versus gate-voltage curves at low temperature. In addition, temperature dependence measurements reveal that the conductance obeys a variable-range-hopping law up to an unexpectedly high temperature of over 100 K. The symmetric dual-gate configuration of the silicon-on-insulator we use allows us to fully characterize the vertical-bias dependence of the conductance. Comparison to computer simulation of the phosphorus impurity band depth-profile reveals how the spatial variation of the impurity-band energy determines the hopping conduction in transistor structures. We conclude that the emergence of the conductance peak and the high-temperature variable-range hopping originate from the band bending and its change by the gate bias. Moreover, the peak structure is found to be strongly related to the density of states (DOS) of the phosphorus impurity band, suggesting the possibility of performing a novel spectroscopy for the DOS of phosphorus, the dopant of paramount importance in Si technology, through transport experiments.

# I. INTRODUCTION

Hopping[1], or tunneling, via dopant atoms in semiconductors has been one of the central issues of semiconductor physics because of its importance for electronic devices and for fundamental phenomena such as the metal-insulator transition (MIT) in disordered systems. Such a dopant-mediated tunneling has been extensively investigated,[1,2] but most of the research has addressed the conduction in bulk semiconductors in which the energy of the impurity band relative to the Fermi energy is only a function of temperature and magnetic field if any. Putting a doped semiconductor into a transistor structure enables us to change the impurity-band energy by using the gate, which allows the hopping conduction to be modulated[3-8] and could provide fruitful information about the impurity band, such as the density of states (DOS). Studies on gate-controlled impurity conduction, however, have concentrated mainly on intentionally contaminated metal-oxide-semiconductor field-effect transistors (MOSFETs) or on the two-dimensional (2D) impurity band of sodium segregated at around $SiO_2$/Si interfaces.[3-6] Very few reports have addressed ordinary bulk or three-dimensional (3D) impurity bands of, for example, phosphorus in silicon (Si:P), which is an important dopant in Si technology.

Gate-induced modulation of the impurity conduction could be investigated using so-called buried-channel MOSFETs,[9] in which the channel and source/drain have the same conduction type, n-type for example. This is because, in this type of device, the impurity band can be located close to the Fermi level of the source terminal of the MOSFET. Buried-channel MOSFETs, however, have not yet been applied for this purpose, though their low-temperature behavior was intensively investigated in the 1970s and 1980s from the viewpoint of their application to low-temperature electronics.[9-14] Those studies were limited only to lightly doped Si in which the tunneling via dopants is not a dominant process for conduction because of negligibly small overlapping of the wave function between dopant electrons.

Besides the field of impurity conduction, investigating the control of the tunneling process in Si:P is important for a certain class of silicon-based quantum information processing[15,16] in which charge manipulation by tunneling via P donors is a critical issue.[17-20] Research on electron transport in Si:P has thus been recently reactivated.[21,22]

This paper reports the characterization of conductance in P-doped buried-channel MOSFETs with doping concentrations higher than those used in previous studies. We find a distinct peak structure to emerge in the conductance vs. gate-voltage curves, and its origin is identified as being a reflection of the DOS of Si:P. In particular, by exploiting a dual-gate structure with front and back gates, we show that the spatial variation of the impurity band energy, which has not been addressed in detail, is decisively important for understanding hopping conduction by the 3D impurity band (like Si:P) in transistors.

## II. DEVICE FABRICATION, CONDUCTANCE MEASUREMENTS AND POTENTIAL-PROFILE SIMULATION

Buried-channel MOSFETs with n-type impurities in both the channel and source/drain regions were fabricated on (100) silicon-on-insulator (SOI) wafers. SOI had been considered an exotic material in the past, but is now recognized as an important one both for advanced Si devices and for a basic understanding of transport[23] in Si owing to its useful dual (front and back) gates, each of which can work equivalently.

The type of SOI used was "SIMOX" (separation-by-implanted oxygen),[24] which was made from boron-doped Si wafers with a doping concentration of the order of $10^{14}$ cm$^{-2}$. The wafers underwent a thermal process at 1350ºC for the buried-oxide formation. Thus, considering the solubility of boron in Si and SiO$_2$, the SOI layer can be regarded as intrinsic Si. This high temperature process also improves the flatness of the back interface dramatically.[25]

P atoms were introduced into this intrinsic SOI layer by ion implantation. Thus, the resultant P-doped SOI is uncompensated. The implantation energy ranged from 40 – 55 keV for a process split, but all the samples were exposed to a thermal activation process at 1000ºC, which is high enough to reach thermal equilibrium in terms of P-atom diffusion in the SOI layer and P-atom exchange between the $SiO_2$/Si interfaces and the SOI layer.[26] We therefore expected that the P-atom distribution would be uniform in the SOI layer. This uniform distribution, which was confirmed by simulation, is one of the merits of using SOI MOSFETs rather than bulk Si MOSFETs and makes data analysis and interpretation straightforward. The final doping concentration ($N_d$) in the SOI layer was evaluated, using eq.(1) in ref. 27, from the deviation of the room-temperature threshold voltage from that of a MOSFET of an unimplanted channel. We observed a considerably large amount of dose loss (80 – 90 %) because of the segregation of P atoms at the interfaces. However, such segregated P atoms are electrically inactive.[28,29] Notice that the $N_d$'s may have about 10% uncertainty, which is mainly due to the uncertainty of the front-gate oxide thickness. A detailed description of the fabrication flow can be found in ref. 27.

We investigated 29-nm-thick wide-area SOI MOSFETs (Fig. 1) with five different values of $N_d$ between 1.2 and 37 x $10^{17}$ $cm^{-3}$. The main focus was 8.7 x $10^{17}$ $cm^{-3}$. This concentration is higher than those previously investigated using bulk Si MOSFETs, which was ~ 1 x $10^{17}$ $cm^{-3}$ or lower,[9-14] but still far lower than the critical concentration $N_c$ (~ 4 x $10^{18}$ $cm^{-3}$) for MIT in uncompensated Si:P.[30]

Two-terminal conductance measurements were performed at temperatures between 10 and 295 K with a zero magnetic field. The source/drain voltage was kept at 30 mV. The temperature was calibrated with a silicon diode temperature sensor. The contact resistance was low (~ 10 Ω or lower) even at low temperatures, which was confirmed using test devices of

metal/n$^{++}$-Si/metal structures, and therefore not an obstacle. The measurement data were reproducible. The conductance characteristics were stable and did not change with time. Also, there were no noticeable differences in conductance characteristics among samples with the same fabrication-process parameters.

To understand the influence of the band bending on the conductance, we calculated the potential depth-profile self-consistently.[31] P donors were treated as a sheet of positive charges uniformly distributed over the entire SOI layer, rather than randomly distributed point charges. We assumed that a P donor can capture only one electron and then becomes neutralized. The energy distribution of the impurity band was described by a $\delta$-function with an ionization energy of 44 meV.[32] This approximation is acceptable for a qualitative argument because, as we will see, the variation of the band energy due to the gate bias is much larger than the expected impurity band width. The doping concentration and thicknesses for the buried oxide and SOI layer were set to the experimental values. The front-gate oxide was adjusted to 95 nm so that the threshold data [Fig. 2(c)] fit the experimental data. This thickness is still within the margin of experimental error. There were no other fitting parameters.

## III. RESULTS AND DISCUSSION

### A. MOSFET with $N_d = 8.7 \times 10^{17}$ cm$^{-3}$

Figures 2(a) shows the conductance $G$, which is the drain current divided by the constant drain voltage of 30 mV, as a function of the front-gate voltage $V_{fg}$ using the back-gate voltage $V_{bg}$ as a parameter. The measurement temperature was 12 K. One can see a pronounced peak structure, which is followed by a sharp rise of the conductance. The peak structure is modulated, or even removed altogether by changing $V_{bg}$. Figure 2(b) shows a gray-scale plot of the conductance in the $V_{fg} - V_{bg}$ plane for the same measurement as Fig. 2(a). Points A, B, and C correspond to those in Fig. 2(a). Point B represents the maximum conductance, while point A represents the minimum conductance. In Fig. 2(b), the gray scale was cycled twice in the displayed conductance window, $10^{-10} - 10^{-4}$ S, in order to accentuate the sharp rising part of the conductance by the white curve.

We first show that this sharp rising is due to the surface channel turning on. Figure 2(c) compares this sharp rising with the threshold voltage of the surface channel calculated by simulation. The experimental data represent the voltage points at which the conductance became $10^{-8}$ S, which is the white curve in Fig. 2(b). For the simulation, the surface-channel threshold voltage was defined as the voltage at which the Fermi level reaches the ground state of the surface channel. The temperature for the simulation was set to 11 K. One can see that the simulated line traces the experimental data well, demonstrating that the sharp rising originates from the surface channel opening. Notice that the choice of the experimental conductance level for the threshold voltage is somewhat arbitrary, but the results do not change significantly because of the sharp threshold characteristics.

Then, the conductance in the left and lower side of the surface-channel threshold can be ascribed to the conduction flowing in the bulk or buried SOI region (We will therefore call this the buried channel), and this is indeed the region where the unusual peak structure is observed.

The voltage conditions for the appearance of the buried- and surface-channel are well explained by considering the P-donor positive charges as follows. First off, the voltage points for the conduction to start are characterized by a slope [the white straight line in Fig. 2(b)] irrespective of the conduction channel; buried or surface. This slope, which was estimated to be -4.1, is determined by the ratio of the front- and back-gate capacitances, i.e., the ratio of the front- and back-gate oxide thicknesses, which are 90 and 380 nm, respectively. This simple explanation is possible because, at these starting points for conduction, nearly all the donors are ionized irrespective of the back gate bias. However, this is not the case for the surface-channel threshold. If $V_{bg}$ is not too high either positively or negatively, the impurity band should reach the Fermi level at a certain buried region before turning the surface channel on because of the U-shaped potential profile, which is created by the P-donor positive charges. This means that some portion of the P donors is neutralized before the surface channel turns on. This leads to the positive shift of the surface-channel threshold voltage. The effect becomes maximum when $V_{bg} \sim 0$ V, which is nearly the flat-band condition. At such a point, the threshold voltage appears at around $V_{fg} = 0$ V, which is basically the same as that for undoped MOSFETs. This is because nearly all donors are neutralized under the flat-band condition. The buried-channel conduction is observed right in the region sandwiched between the current-starting line [the line parallel to the white straight line in Fig. 2(b)] and the surface-channel threshold curve. We should point out that the surface-channel threshold-voltage curve (and even the entire conductance structure) has a mirror symmetry with respect to $V_{bg} = c^{-1}V_{fg}$, with $c$ of the gate-capacitance ratio, 4.1, supporting a uniform distribution of P donors in the SOI layer.

The above results indicate that the overall conductance structure in the $V_{fg} - V_{bg}$ plane is governed by the variation of the number of P donor positive charges due to the gate bias. However, they do not account for the conductance dip that appears right before the surface-channel opening. We thus investigated the temperature dependence of the conductance in order to explore the conduction mechanisms of the buried channel. Figure 3 shows the results for the back-gate voltage $V_{bg} = -8$ V, which includes the conductance maximum point (point B). Figure 4(a) shows the conductance as a function of $T^{-1/4}$ at the peak or at point B. The curve is fitted well by a straight line for temperatures up to about 100 K. This suggests that the conductance is dominated by variable-range hopping (VRH). (The conductance as a function of the inverse temperature for this device can be found in curve II in Fig. 9.)

In general, VRH conduction has the form $G = G_0(T_0/T)^s \exp(-\xi_c)$, with $\xi_c = (T_0/T)^p$, where $G_0$ is a constant independent of temperature, and $s$ is a small value, which could vary depending on detailed models of VRH. The $\xi_c$ is the percolation threshold and $T_0$ is the characteristic temperature. The exponent $p$ depends on the dimension $d$ of space in which hopping takes place and by the strength of the Coulomb interaction between electrons. 3D and 2D hopping without the Coulomb interaction leads to $p = 1/4$ and $1/3$, respectively, while that with Coulomb interaction leads to $p \sim 1/2$ irrespective of $d$, which is also known as Efros-Shklovskii (ES) VRH.[33] Determining the exponent $p$ requires a careful treatment of the pre-exponential factor. Although the pre-exponential factor is weakly dependent on the temperature, it indeed affects the determination of the exponent. We thus evaluated the fitting error as a function of $p$ using $s$ as a parameter. For this error estimation, we calculated the sum of $(\ln G_{mes} - \ln G_{fit})^2$ for temperatures between 13 – 50 K. Here, $G_{mes}$ is the measured value and $G_{fit}$ is the theoretical one. Figure 4(b) shows the results. In the figure, we plot, as bold lines, the results for $s$'s close to the one obtained from percolation theory,[1] which gives $s = 0.2 – 0.3$,[34] regardless of the dimension. One can see that the $p$ giving minimum error is around $0.25 – 0.27$. Therefore, the most plausible value for $p$ is 1/4. In other words, the conductance obeys the Mott law or 3D VRH without Coulomb interaction.

Provided the Mott law stands, the characteristic temperature $T_0$ is estimated to be 1.4 x $10^6$ K. This high value of $T_0$ is consistent with the previous data for bulk Si:P with a doping concentration far lower than $N_c$.[35-37]

The mean hopping length $\bar{R}$ can be estimated by $0.3\alpha\xi_c$,[38] where $\alpha$ is the localization radius in hopping conduction. As we have mentioned, the present $N_d$ (= 8.7 x 10$^{17}$ cm$^{-3}$) is very low deep in the insulating side of MIT. For such a diluted doping, $\alpha$ will be close to or only slightly larger than the characteristic decay length $a_0$ of the wave function for isolated donors.[36] This is expressed as $a_0 = \sqrt[3]{a^2 b}$, where $a = h/2\pi\sqrt{2m_t E_d}$ and $b = h/2\pi\sqrt{2m_l E_d}$, with $h$ denoting the Plank's constant, $E_d$ the ionization energy, $m_t$ and $m_l$ the transverse and longitudinal effective masses in the Si conduction band, respectively. Provided $E_d$ = 44 meV,[32] $a_0$ comes to 1.62 nm. If we assume $\alpha = a_0$, then $\bar{R}$ comes to 9 and 5 nm at 10 and 100 K, respectively. These are the same order as the mean distance ($\sim N_d^{-1/3}$) of the P donors, 10 nm, and thus reasonable. Also, the above value for $\bar{R}$ is smaller than the SOI layer thickness, 29 nm, which is consistent with 3D conduction.

The mean hopping energy $\bar{E}$ can be estimated using $0.39\xi_c kT$,[38] where $k$ is the Boltzmann constant. This results in $\bar{E}$ = 37 meV at 100 K. This large value directly reflects the fact that VRH was observed at high temperatures. Considering the impurity band width of Si:P with the present $N_d$, which is around 5 meV,[39-41] this large $\bar{E}$ may appear surprising. In VRH, the allowed energy range can be defined by $\bar{E}$. Therefore, when the impurity-band width is narrower than $\bar{E}$, we should not observe VRH. Instead, we should observe nearest-neighbor hopping, which obeys a thermal activation law with a single activation energy. As will be shown, however, gate-bias-induced band bending in the SOI layer widens the effective band width, which allows electrons to jump to high energy states, making VRH possible. This is what makes the Si:P in transistors distinctive from bulk Si:P.

Figure 5 shows the $T^{-1/4}$ plots for various $V_{fg}$ and $V_{bg}$. As shown in the figure, the data can still be fitted with $p = 1/4$ well. We should however point out that the fitting error under the assumption of $p = 1/4$ was found to increase as $V_{fg}$ and $V_{bg}$ deviated from the peak position (point B) to the negative-voltage side and that the optimum value for $p$ shifts from 1/4 towards 1/3 there. This suggests 3D – 2D transition of hopping dimension, but in this paper we do not analyze this point in any detail; this will be discussed elsewhere.

In spite of this uncertainty of the VRH dimension in the low $V_{fg} - V_{bg}$ regions, the fitting analysis showed that we could exclude the possibility of the ES VRH, which should give $p \sim 1/2$. The ES VRH is not plausible for the following reason as well. In the ES VRH, the conductance takes the form of $G = G_0' \exp[-(T_0'/T)^{1/2}]$, with $T_0' = 7.2e^2/(\alpha k \varepsilon)$,[38] where $e$ and $\varepsilon$ are the unit charge and the permittivity of silicon, respectively. If, for example, we fit the data for the maximum conductance point (point B in Fig. 2) with this form (though the fitting error was much larger than the case for $p = 1/4$), we obtained $T_0' = 2.1 \times 10^3$ K. From this value, the localization length comes to 60 nm using the expression for $T_0'$, which is unrealistically larger than the expected value of 1.62 nm. From the above data and analysis, we concluded that the buried-channel conductance is governed by VRH without Coulomb interaction, especially by 3D VRH around the peak region (point B).

This conclusion led us to perform further analysis using the expression for the characteristic temperature, which is $T_0 = \dfrac{\beta}{kg(E_f)\alpha^3}$, where $\beta (= 10.0)$[38] is a constant and $g(E_f)$ is the DOS of the impurity band at the Fermi level. Figure 6 shows $T_0$ and the resultant $g(E_f)$ as a function of $V_{fg}$ for $V_{bg} = -8$ V. $T_0$ was derived from the slopes of the VRH plots like those shown in Figs. 4(a) and 5(a). One can see that the $g - V_{fg}$ curve has a maximum and becomes smaller when the conductance becomes smaller. This suggests that the conductance curve is related to the energy distribution of the DOS of the impurity band.

The value of $g(E_f)$ at the peak is about 2.3 x $10^{19}$ cm$^{-3}$eV$^{-1}$. Upon a flat DOS around $E_f$, and using the relation $N_d = g(E_f)W$, the band width $W$ comes to 38 meV. This is again larger than the expected width of the impurity band at the present $N_d$. In turn, this wide band means that the obtained $g(E_f)$ is smaller than that for bulk Si:P. In other words, only some portion of P donors in the SOI layer contributes to the hopping conduction. The mean hopping energy $\bar{E}$ at 12 K is 7.9 meV, and the density of the P donors contributing to the peak conduction at 12 K thus comes to $\bar{E}\ g(E_f) \sim 2$ x $10^{17}$ cm$^{-3}$. This is slightly smaller than the present $N_d$ (= 8.7 x $10^{17}$ cm$^{-3}$).

To qualitatively explain the above experimental results, we show in Fig. 7 the calculated depth-profile of the impurity band at 11 K. Fig. 7(b) shows the case for $V_{bg}$ = - 8 V. The horizontal axis is the SOI depth with the front/back interface corresponding to 0/29 nm. The vertical axis is the impurity level $E_d$ with respect to the Fermi level $E_f$. We also indicate the allowed band for VRH, $0.39\xi_c kT$, which is 7.2 meV at 11 K, by the dashed lines. Six curves labled B1 – B6 are the calculated data for voltage points indicated in Fig. 7(a). In B1, even the minimum energy point is above the $E_f$. In B2, a point is about to reach $E_f$, where the donors become neutralized. For this point and for larger $V_{fg}$'s (B2 – B6), the band energy is pinned at the back-interface side. That is, the band profile at the back-interface side is unchanged in B2 – B6 because positive charges induced by $V_{fg}$ are compensated for by means of the neutralization of the donors. For B3 and B4, such a neutralized region extends towards the front-interface side. Even in the neutralized region, the band is not perfectly flat because a small fraction of donors is kept ionized due to finite temperature. In B5, we see the flat band at the front-interface side. At this point, there remain very few donors to be neutralized. Therefore, a further increase in $V_{fg}$ causes a sudden decrease of the band energy at the front-interface side as shown in curve B6.

The VRH conduction should take place most frequently in the region where the condition $E_d = E_f$ is satisfied. Curves B3 and B4 have two such points, while B2, B5, and B6 have just one, and B1 has no such point. *This is the origin of the conductance peak.* Namely, by changing the voltage points from B1 to B6, we actually scan the impurity-band energy from a very low to a very high value at the front-interface side, which is in effect spectroscopy of the DOS.

The calculated curves also indicate that the VRH mostly occurs at a certain depth in the SOI layer, i.e., at the points where $E_d$ and $E_f$ cross. The real situation is however a bit more complicated because of the finite temperature, which widens the allowed band defined by the mean hopping energy $\bar{E}$. Looking at curves B3 and B4, the extension of the SOI region within $\bar{E}$ is on the order of 10 nm for each crossing point. Even if this broadening occurs, we can still say that the VRH current flows non-uniformly in depth; more current around the crossing points and less current near the interface regions. This is why the donor density that contributes to the conductance is smaller than $N_d$. From Fig. 7(b), we also understand that VRH can be observed even at very high temperatures. This is because we can still find empty states for electrons to hop to, due to the large band bending.

Figure 7(c) shows the case for $V_{bg} = 1$ V, which corresponds to the experimental conductance curve for $V_{bg} = 0$ V in Fig. 2(a). (In order to make our claim clear, we show the calculation results for $V_{bg} = 1$ V rather than for $V_{bg} = 0$ V. Because the substrate of the SOI wafer used was p-type Si, the exact flat band condition at the back interface appears at $V_{bg} = 1$ V. The experimental $G - V_{fg}$ curve was not significantly changed between $V_{bg} = 1$ and 0 V.) Compared to the case for $V_{bg} = -8$ V, because of the flat band at the back interface, only one crossing point can be seen, which makes the SOI region within the allowable window *thinner*. This is why the peak conductance for $V_{bg} = 0$ V was smaller than that for $V_{bg} = -8$ V [Fig. 2(a)]. It also explains why the conductance minimum appears at around $V_{fg} = V_{bg} = 0$ V. At this voltage condition, the band is nearly flat and beneath the $E_f$ line. Therefore, most donors are neutralized, making it difficult to find empty donor states for hopping.

Figure 7(d) shows the case when we apply a large bias either at the front or back gate. As one can see, the potential is inclined very strongly and the entire band is nearly out of the allowed range, resulting in the disappearance of hopping conduction.

In summary for $N_d = 8.7 \times 10^{17}$ cm$^{-3}$, the overall structure of the conductance, e.g., the surface-channel threshold, in the $V_{fg} - V_{bg}$ plane was well explained by considering the P-donor positive charges. We observed a peak structure in the conductance curve, which was found to obey the Mott law. The estimated values for the characteristic temperature and the mean hopping length were reasonable, but the mean hopping energy was unexpectedly large due to a high-temperature VRH. This cannot be explained by simple VRH theory, but seems to be explainable if we consider the large band bending, indicating the importance of the spatial variation of the impurity-band energy. The potential-profile simulation can explain the maximum and minimum conductances that appear in the negative $V_{fg} - V_{bg}$ region and at around $V_{fg} = V_{bg} = 0$ V, respectively. It also strongly suggests that the peak structure originates from the DOS of Si:P.

### B. Dependence on doping concentration

We observed similar conductance peaks for MOSFETs with $N_d = 2.1$ and $5.0 \times 10^{17}$ cm$^{-3}$ (data not shown). We found that the peak conductance decreased nearly exponentially as a function of $N_d^{-1/3}$, i.e., the mean donor distance. This reasonably agrees with the nature of hopping because the magnitude of the conductance is dominated by the overlap integral of the wave functions. Because of this rapid decrease of the conductance, we did not observe the peak structure for MOSFETs with $N_d$ = 1.2 x $10^{17}$ cm$^{-3}$, the lowest concentration we examined. (The buried channel conductance was lower than the detectable limit at low temperatures.) In turn, higher doping concentration ($N_d$ = 3.7 x $10^{18}$ cm$^{-3}$) also resulted in the disappearance of the peak structure. Figure 8 shows the temperature dependence of the conductance for these two MOSFETs for $V_{bg}$ = 0 V, and Fig. 9(a) shows the conductance as a function of the inverse temperature for a fixed $V_{fg}$ with $V_{bg}$ = 0 V. For comparison, the data for the main sample with $N_d$ = 8.7 x $10^{17}$ cm$^{-3}$ is also plotted in Fig. 9(a).

Figure 8(a) shows the data for $N_d$ = 1.2 x $10^{17}$ cm$^{-3}$. As in the case for the main sample with $N_d$ = 8.7 x $10^{17}$ cm$^{-3}$, the current starts to flow in the negative $V_{fg}$ region at high temperatures, but with a smaller negative $V_{fg}$ due to the lower $N_d$. As the temperature decreases, the conductance rapidly decreases in this negative $V_{fg}$ region and becomes lower than the detectable limit at 25 K or lower. This is because of the strong carrier freezeout: trapped electrons do not contribute to the conduction because of a small overlap integral of the wave functions. As one can see in Fig. 9(a), for this lowest concentration (curve I), the conductance obeys a thermal activation law, suggesting that the conductance is due to the conduction-band electrons. A more accurate description for this concentration is the following. For all $V_{fg}$, the temperature dependence was found to be fitted much better by the activation law rather than by the VRH law, i.e., by $T^p$ with $p$ = 1/2, 1/3, or 1/4. Then, as shown in Fig. 9(b), the activation energy $E$ was found to decrease monotonically with the increased $V_{fg}$, but to have an inflection point at a certain $V_{fg}$ (~ - 0.65 V). The activation energy at this inflection point was 45 meV. This indicates that a weak Fermi level pinning occurs at the impurity level, and that the conduction is governed by the conduction-band electrons under such conditions.

For the sample with the highest concentration, $N_d = 3.7 \times 10^{18}$ cm$^{-3}$, which is close to $N_c$, we observed the conductance to be weakly dependent on the temperature [Fig. 8(b)]. This sample will therefore still be in the insulating side of MIT. The characteristics can be fitted by VRH for lower temperature side, and $T_0$ was estimated to be of the order of $10^3$. If we compare this value for $T_0$ to the case for bulk Si:P, then $1 - N_d/N_c$ comes to around 0.2,[32] which results in $N_d \sim 3 \times 10^{18}$ cm$^{-3}$. This agrees with the doping concentration we evaluated using the room-temperature threshold voltage. At such a high concentration, because of the strong coupling between donor electrons, the width of the bands, especially, that of the upper Hubbard band become significantly larger.[35,36] This could destroy the gap between the impurity band (or the lower Hubbard band) and the mobility edge, leading to the disappearance of the conductance peak. The disappearance of the conductance dip in this high-concentration sample in turn supports the idea that the dip for $N_d = 8.7 \times 10^{17}$ cm$^{-3}$ is related to the energy gap between the impurity band and the mobility edge.

From the above-mentioned $N_d$ dependence and the analysis of the main $N_d$, we conclude that the observed peak structure originates from the DOS of Si:P.

C. Discussion

We here discuss the application of the present results to the derivation of fundamental quantities for hopping. We first emphasize that the present characterization of the conductance data owes largely to the use of SOI as a base material. As indicated in a previous study,[26] a high-temperature thermal process makes it possible to obtain a uniform distribution of dopants over the entire SOI layer, which makes the analysis of the conductance data easier. This uniform distribution is difficult to achieve if we use bulk MOSFETs because the Si is deeply extended, which prevents the system from achieving thermal equilibrium in terms of the dopant diffusion and dopant exchange between the interface and host Si. The dual gate configuration is another merit of the SOI. The use of the substrate bias in bulk MOSFETs is restrictive because it is asymmetric with respect to the front gate and is basically effective only for negative bias. Therefore, the SOI is a good experimental host to investigate the impurity conduction in transistors.

The present results suggest a possibility of extracting the DOS structure and even its depth profile for Si:P from the conductance data. For example, moving $V_{fg}$ from point A2 to A3 in Fig. 7(c) corresponds to scanning the impurity-band energy at the front interface side. A detailed analysis will make it possible to compare the resultant DOS to the vast amount of accumulated data of optical DOS for Si:P. However, the relationship between the DOS and the conductance curve is indirect because the potential profile changes in a complicated manner with the gate bias as we have shown. Thus, we require a more sophisticated model for the present VRH in order to make quantitative arguments.

Another important implication of the present results arises from the fact that the introduction of a non-zero electric field *perpendicular to* the direction for the hopping current to flow plays a crucial role for the hopping conduction. This parameter is rather new and has not yet been addressed in detail. We claim that this non-zero field made it possible for VRH to occur at such high temperatures. In addition, the band-profile calculation suggested a non-uniform flow of current as a function of depth because of the band bending. It would surely be interesting to study more deeply the conduction under the voltage conditions where the allowed region becomes very *thin*, like the case for curve B2 in Fig. 7(b). In such a case, 2D hopping rather than 3D hopping is expected. We would be able to study, by controlling the effective SOI thickness for conduction using the gate biases, the 3D – 2D transition of hopping in doped semiconductors, like the study[8] for 2D – 1D transition in a 2D electron gas of MOSFET surface channels. Such gate-bias-controlled experiments may provide a new reliable way of extracting the localization radius for hopping conduction in Si:P, similar to the method[42] proposed by Knotek et al. for amorphous semiconductors, but without the need for preparing samples of different film thicknesses, which can cause many uncontrollable factors to be introduced.

## IV. CONCLUSIONS

Transport measurements have been performed for P-doped buried-channel SOI MOSFETs in a temperature range between 10 – 295 K. For doping concentration of the order of $10^{17}$ cm$^{-3}$, a clear peak was observed in the conductance versus gate-voltage curves at low temperatures. It was also found that the conductance can be explained by variable-range hopping up to high temperatures of around 100 K. The calculations of the impurity-band profile showed that the gate-bias-induced energy variation of the P impurity band plays a crucial role for these phenomena.

**Acknowledgments**

We thank A. Fujiwara, M. Uematsu, H. Kageshima, and Y. Sato of NTT for their helpful discussions. We also thank H. Namatsu, M. Nagase, K. Yamazaki, Y. Watanabe, J. Hayashi, T. Yamaguchi, and K. Inokuma of NTT for their collaboration in device fabrication. This work was partly supported by MEXT KAKENHI (16206038 and 17201029).

**References**

1. B. I. Shklovskii and A. L. Efros, *Electronic properties of doped semiconductors*, Springer-Verlag, Berlin, 1984.

2. M. Pollak and B. I. Shklovskii, *Hopping transport in Solids, North-Holland*, Amsterdam, 1991.

3. A. Hartstein and A. B. Fowler, Phys. Rev. Lett. **34**, 1435 (1975).

4. A. B. Fowler, G. L. Timp, J. J. Wainer, and R. A. Webb, Phys. Rev. Lett. **57**, 138 (1986).

5. E. Glaser and B. D. McCombe, Phys. Rev. **B37**, 10769 (1988).

6. T. Ferrus, R. George, C. H. W. Barnes, N. Lumpkin, D. J. Paul, and M. Pepper, Phys. Rev. **B73**, 041304R (2006).

7. M. Pepper, Philos. Mag. 37B, 187 (1978).

8. A. B. Fowler, A. Hartstein, R. A. Webb, Phys. Rev. Lett. **48**, 196 (1982).

9. F. H. Gaensslen and R. C. Jaeger: Solid-State Electron **22,** 423 (1979).

10. R. C. Jaeger and F. H. Gaensslen: IEEE Trans. Electron Devices **26**, 501 (1979).

11. R. C. Jaeger and F. H. Gaensslen: IEEE Trans. Electron Devices **27,** 914 (1980).

12. S. K. Tewksbury, M. R. Biazzo, T. R. Harrison, T. L. Lindstrom, D. N. Tennant and G. Storz: IEEE Trans. Electron Devices **26,** 501 (1979).

13. S. K. Tewksbury, M. R. Biazzo and T. L. Lindstrom: IEEE Trans. Electron Devices **32** (1985) 67.

14. R. A. Wilcox, J. Chang and C. R. Viswanathan: IEEE Trans. Electron Devices **36,** 1440 (1989).

15. B. E. Kane, Nature **393,** 133 (1998).

16. L. C. L. Hollenberg, A. S. Dzurak, C. Wellard, A. R. Hamilton, D. J. Reilly, G. J. Milburn and R. G. Clark, Phys. Rev. **B69**, 113301 (2004).

17  M. Kettle, H.-S. Goan, S. C. Smith, C. J. Wellard, L. C. L. Hollenberg and C. I. Pakes, Phys. Rev. **B68**, 075317 (2004).

18  G. D. J. Smit, S. Rogge, J. Caro and T. M. Klapwijk: Phys. Rev. B**68**, 193302 (2003).

19  A. S. Martins, R. B. Capaz and Belita Koiller, Phys. Rev. **B69**, 085320 (2004).

20  M. J. Caldero´n, Belita Koiller, Xuedong Hu, and S. Das Sarma1, Phys. Rev. Lett. **96,** 096802 (2006).

21  T.-C. Shen, J. S. Kline, T. Schenkel, S. J. Robinson, J.-Y. Ji, C. Yang and R.-R. Du, J. R. Tucker, J. Vac. Sci. technol. **B22**, 3182 (2004).

22  T. M. Buehler et al., Appl. Phys. Letts. **88**, 192101 (2006).

23  K. Takashina, Y. Ono, A. Fujiwara, Y. Takahashi, and Y. Hirayama, Phys. Rev. Lett. **96,** 236081 (2006).

24  K. Izumi, M. Doken, and H. Ariyoshi, Electron. Lett. **14**, 593 (1978).

25  Y. Takahashi, A. Fujiwara, M. Nagase, H. Namatsu, K. Kurihara, K. Iwadate, and K. Murase, Int. J. Electronics **186**, 605 (1999).

26  Y. Sato, K. Imai and E. Arai, J. Electrochem. Soc. **142**, 660 (1995).

27  Y. Ono, K. Nishiguchi, H. Inokawa, S. Horiguchi and Y. Takahashi, Jpn. J. Appl. Phys. **44**, 2588 (2005).

28  H. R. Soleimani, J. Electrochem. Soc. **141**, 2182 (1994).

29  Y. Sato, M. Watanabe and K. Imai, J. Electrochem. Soc. **140,** 2679 (1993).

30  T. F. Rosenbaum, K. Andres, G. A. Thomas, R. N. Bhatt, Phys. Rev. Lett. **45**, 1723 (1980).

31  S. Horiguchi, A. Fujiwara, H. Inokawa and Y. Takahashi, Jpn. J. Appl. Phys. **43,** 2036 (2004).

32  S. M. Sze and J. C. Irvin, Solid-State Electron. **11** 599 (1968).

33  A. L. Efros and B. I. Shklovskii, J. Phys. **C8**, L49 (1975).

34  The exponent *s* can be expressed by $(2 - \upsilon)/d$, where $\upsilon$ is the critical exponent for the correlation radius. For $d = 2$ and 3, $\upsilon \sim 1.3$ and 0.9, respectively, which gives $s = 0.23$ and 0.275 for $d = 2$ and 3, respectively.

35  W. Sasaki, Philos. Mag. B52, 427 (1985).

36  W. Sasaki, Y. Nisho, and K. Kajita, in *Disordered Semiconductors* (Plenum Press, New York, 1987) p. 37.

37  Fig. 6 in chapter 1 of ref.2.

38  In order to evaluate the mean hopping length, the mean hopping energy and the characteristic temperatures from the experimental value of the percolation threshold $\xi_c$, it is necessary to calculate the dimensionless critical concentration *n* based on the percolation theory. We calculated *n* by considering the anisotropic nature of the donor electrons in Si:P to be 2.6 (unpublished), and used this value for the estimation of the above quantities. However, the deviation of the results from the ones using the conventional isotropic wave functions ($n = 5.3 \pm 0.3$) is small, and thus, this modification does not affect any of our qualitative arguments.

39  M. Kikuchi, J. Phys. Soc. Jpn. **25**, 989 (1965).

40  R. N. Bhatt and T. M. Rice, Phys. Rev. **B23**, 1920 (1981).

41  G. A. Thomas, M. Capizzi, F. DeRosa, R. N. Bhatt and T. M. Rice, Phys. Rev. **B23**, 5472 (1981).

42  M. L. Knotek, M. Pollak, T. M. Donovan, H. Kurtzman, Phys. Rev. Lett. **30**, 853 (1973).

**FIGURE CAPTIONS**

FIG. 1. (color online) Cross-sectional and top views of the MOSFET. The gate length and width are 14 and 10 μm, respectively. The voltages to the front and back gate are denoted by $V_{fg}$ and $V_{bg}$, respectively.

FIG. 2. (color online) Conductance characteristics for a MOSFET with $N_d = 8.7 \times 10^{17}$ cm$^{-3}$, measured at 12 K. Conductance as a function of the front-gate voltage for three back-gate voltages (a), gray-scale image of the conductance in the $V_{fg}$ - $V_{bg}$ plane (b), and experimental (dots) and simulated (line) surface-channel threshold voltage in the $V_{fg}$ - $V_{bg}$ plane (c). In (b), the gray scale was cycled twice in the displayed conductance window, $10^{-10} - 10^{-4}$ S, in order to accentuate the sharp rising part of the conductance by the white curve. In (c), the buried-channel region is also indicated.

FIG. 3. Temperature dependence of $G(V_{fg})$ curve for a MOSFET with $N_d = 8.7 \times 10^{17}$ cm$^{-3}$.

FIG. 4. (color online) Conductance as a function of $T^{-1/4}$ (a) and the fitting error as a function of the exponent $p$ (b), for a MOSFET with $N_d = 8.7 \times 10^{17}$ cm$^{-3}$.

FIG. 5. Conductance as a function of $T^{-1/4}$ for $V_{bg} = -8$ V (a) and 0 V (b).

FIG. 6. The $V_{fg}$ dependence of the characteristic temperature $T_0$ (a) and the DOS at the Fermi level (b).

FIG. 7. (color online) Simulated depth profile of the P impurity level relative to the Fermi level, at 11 K. Voltage points for the calculation are shown in (a). Graphs (b) and (c) correspond to the back-gate voltage of – 8 and 1 V, respectively. Depth profiles in (d) correspond to the back-gate voltages of 10 and – 40 V. Two dashed horizontal lines in each figure (b – d) define the mean hopping energy at 11 K.

FIG. 8. Conductance as a function of the front-gate voltage for MOSFETs with $N_d$ = 1.2 x $10^{17}$ $cm^{-3}$ (a) and 3.7 x $10^{18}$ $cm^{-3}$ (b). For both sets of data, the back-gate voltage was kept at 0 V.

FIG. 9. (a) Conductance as a function of the inverse temperature for MOSFETs with three different $N_d$'s (a) and the activation energy $E$ as a function of $V_{fg}$ for the MOSFET with $N_d$ = 1.2 x $10^{17}$ $cm^{-3}$ (b). In (a), all three curves were obtained with the back-gate voltage kept at 0 V. The front-gate voltages were – 0.65, – 4.0, and – 10 V for curves I – III, respectively.

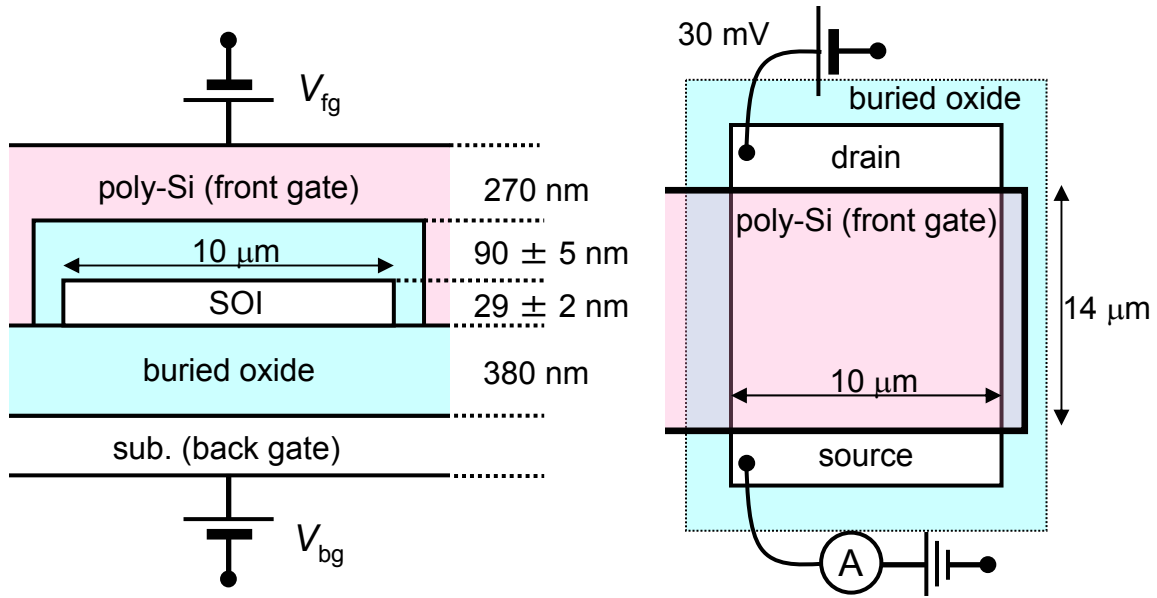

Y.Ono et al., Fig. 1

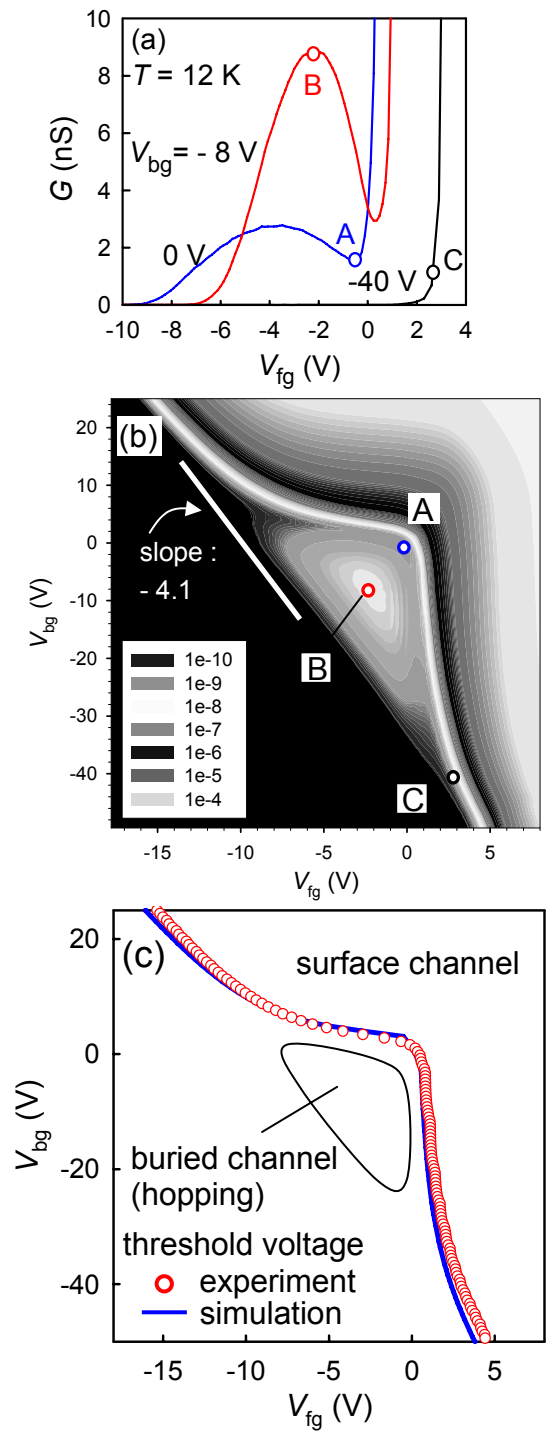

Y.Ono et al., Fig. 2

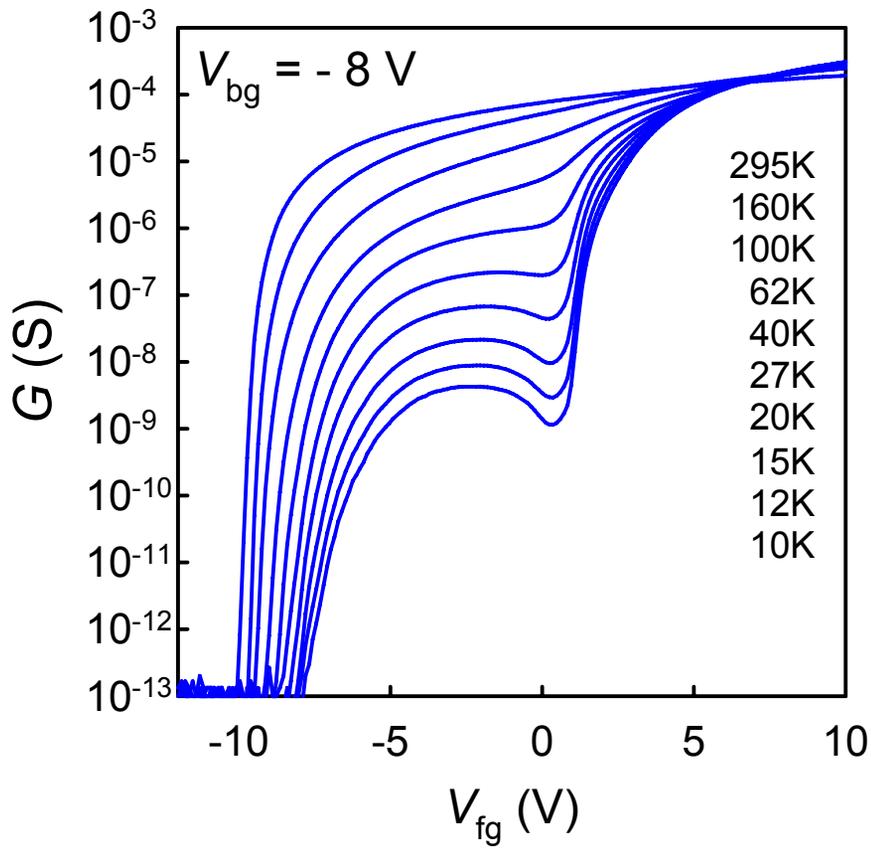

Y.Ono et al., Fig. 3

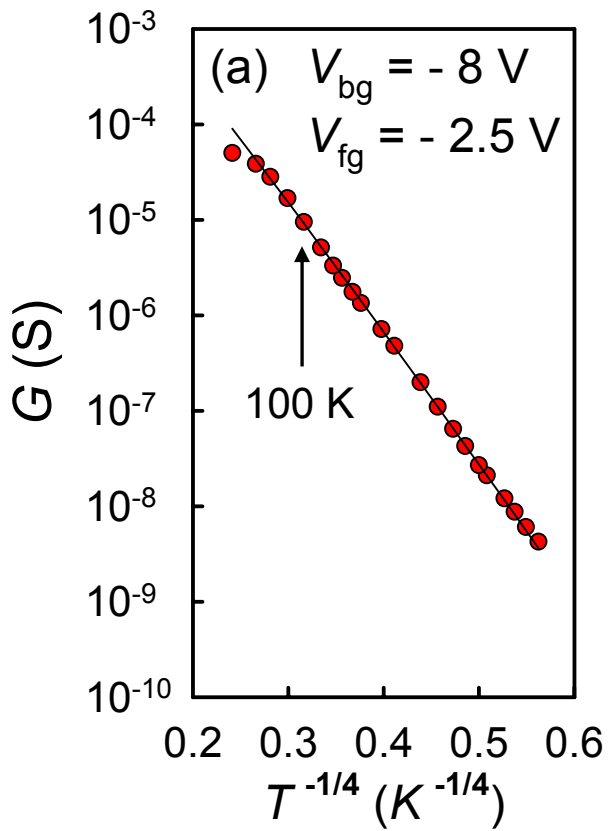
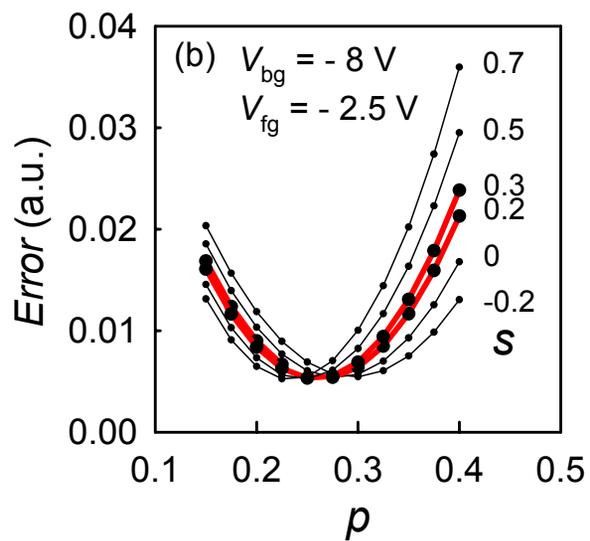

Y.Ono et al., Fig. 4

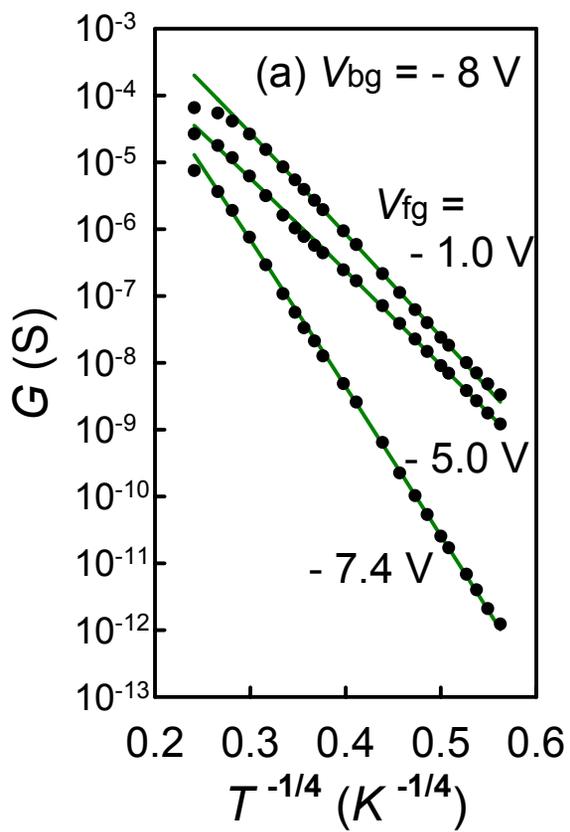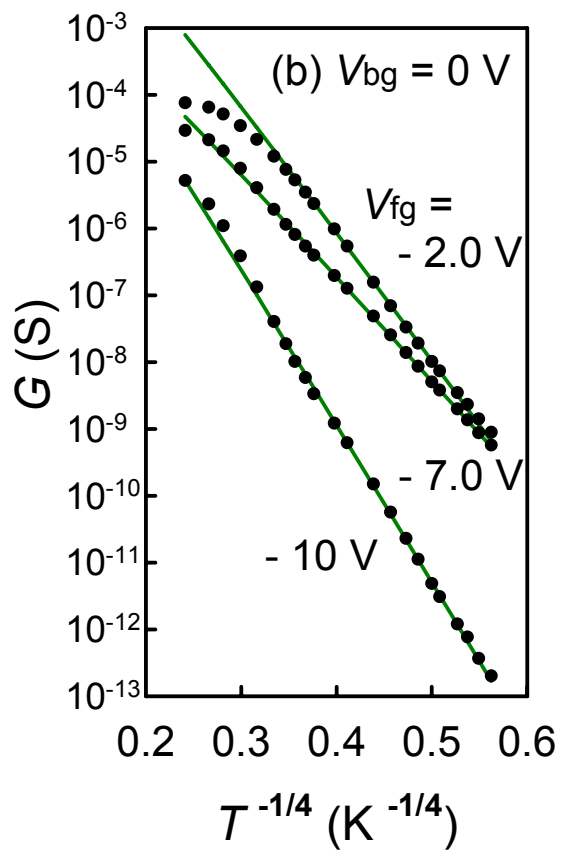

Y.Ono et al., Fig. 5

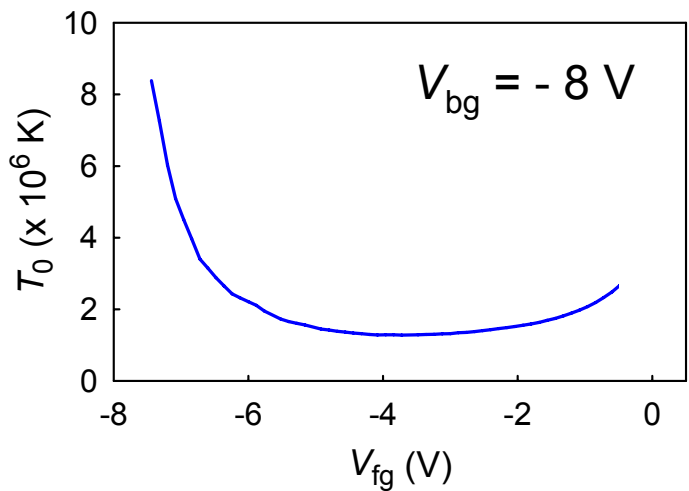
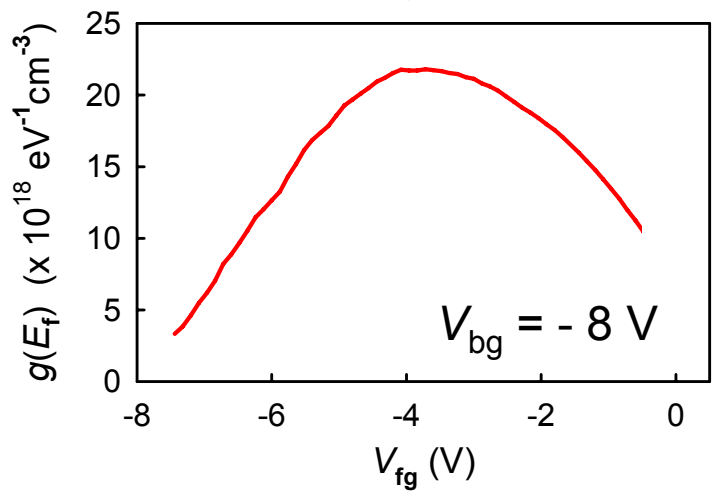

Y.Ono et al., Fig. 6

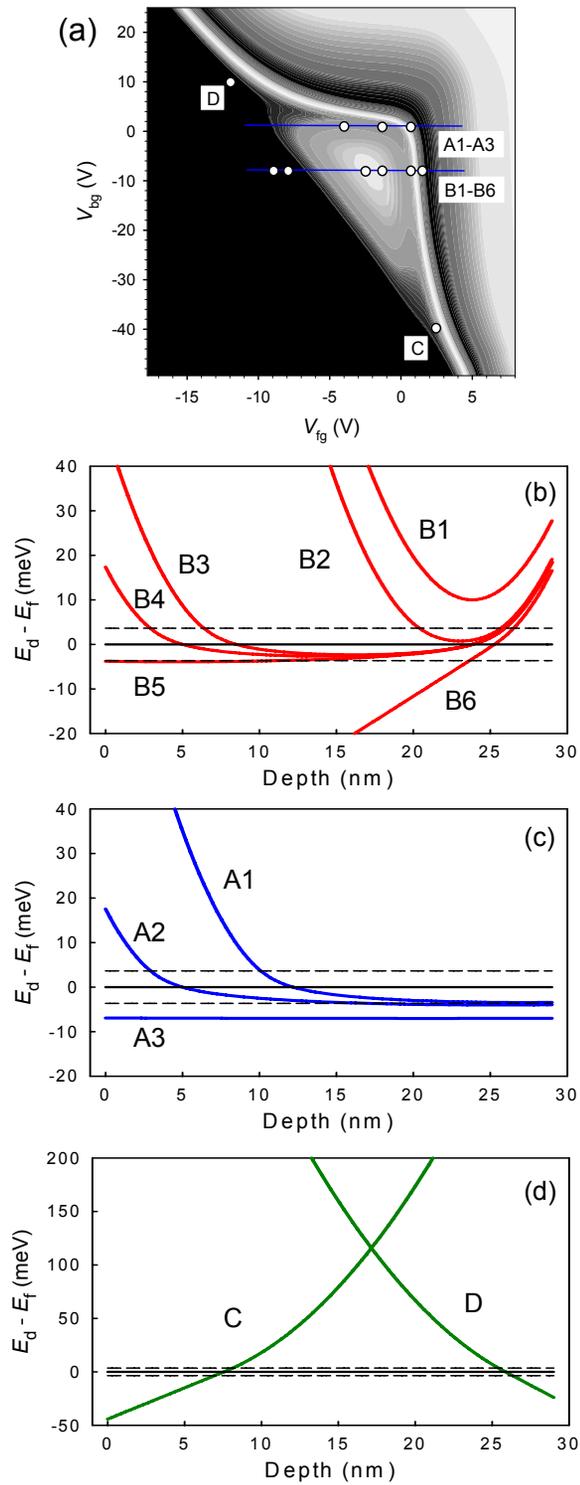

Y.Ono et al., Fig. 7

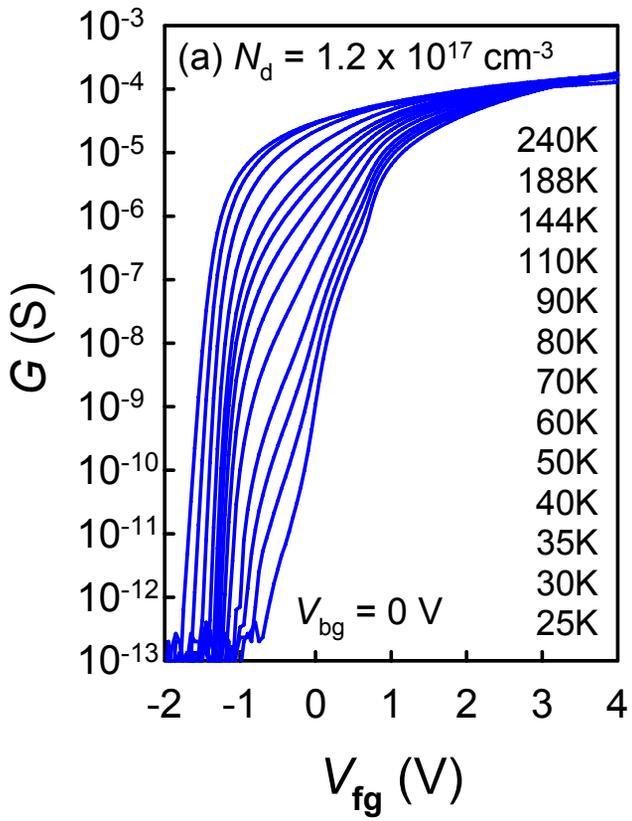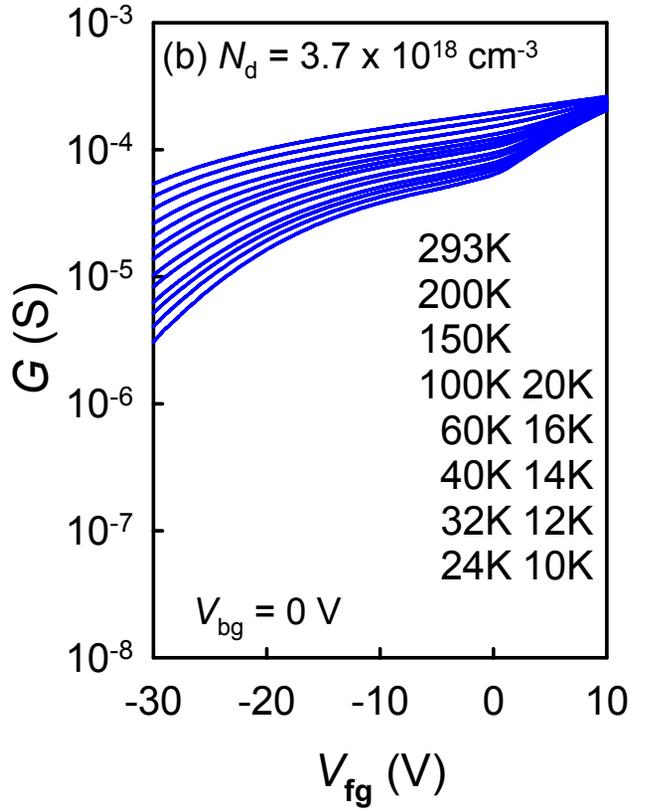

Y.Ono et al., Fig. 8

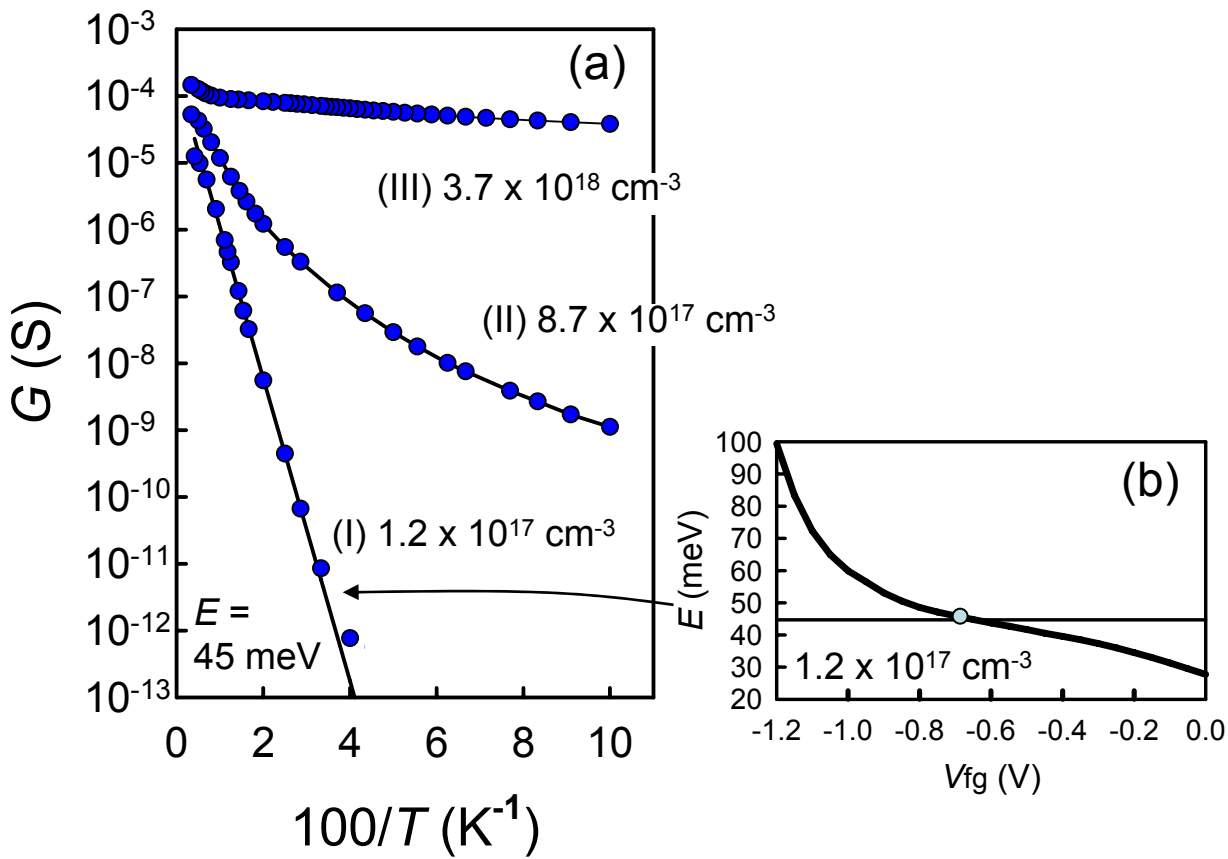

Y.Ono et al., Fig. 9